\documentclass[submission,copyright,creativecommons]{eptcs}
\providecommand{\event}{PLACES 2026} 

\usepackage{iftex}
\usepackage{listings}
\usepackage{xcolor}
\usepackage{verbatim}
\usepackage{todonotes}
\usepackage{enumitem}

\ifpdf
  \usepackage{underscore}         
  \usepackage[T1]{fontenc}        
\else
  \usepackage{breakurl}           
\fi

\definecolor{GrayCodeBlock}{RGB}{241,241,241}
\definecolor{BlackText}{RGB}{110,107,94}
\definecolor{RedTypename}{RGB}{182,86,17}
\definecolor{GreenString}{RGB}{96,172,57}
\definecolor{PurpleKeyword}{RGB}{184,84,212}
\definecolor{GrayComment}{RGB}{170,170,170}
\definecolor{GoldDocumentation}{RGB}{180,165,45}
\lstdefinelanguage{Viper} 
{
    columns=fullflexible,
    keepspaces=true,
    frame=single,
    framesep=0pt,
    framerule=0pt,
    framexleftmargin=4pt,
    framexrightmargin=4pt,
    framextopmargin=5pt,
    framexbottommargin=3pt,
    xleftmargin=4pt,
    xrightmargin=4pt,
    backgroundcolor=\color{GrayCodeBlock},
    basicstyle=\ttfamily\color{BlackText},
    keywords={
        true,false,
        unsafe,async,await,move,
        use,pub,crate,super,mod,
        struct,enum,fn,const,static,let,mut,ref,type,impl,where,as,
        break,continue,if,else,while,for,loop,match,return,yield,in,predicate,field,acc,requires,ensures, subtype, assert
    },
    keywordstyle=\color{PurpleKeyword},
    ndkeywords={
        bool,char,str,
        Self,Option,Some,None,Result,Ok,Err,String,Box,Vec,Rc,Arc,Cell,Ref,HashMap,BTreeMap, NonZero, Nat, int, NonNull, Index,
        macro_rules
    },
    ndkeywordstyle=\color{RedTypename},
    comment=[l][\color{GrayComment}\slshape]{//},
    morecomment=[s][\color{GrayComment}\slshape]{/*}{*/},
    morecomment=[l][\color{GoldDocumentation}\slshape]{///},
    morecomment=[s][\color{GoldDocumentation}\slshape]{/*!}{*/},
    morecomment=[l][\color{GoldDocumentation}\slshape]{//!},
    morecomment=[s][\color{RedTypename}]{\#![}{]},
    morecomment=[s][\color{RedTypename}]{\#[}{]},
    stringstyle=\color{GreenString},
    string=[b]"
}
\lstdefinelanguage{JavaCors} {
    columns=fullflexible,
    keepspaces=true,
    frame=single,
    framesep=0pt,
    framerule=0pt,
    framexleftmargin=4pt,
    framexrightmargin=4pt,
    framextopmargin=5pt,
    framexbottommargin=3pt,
    xleftmargin=4pt,
    xrightmargin=4pt,
    backgroundcolor=\color{GrayCodeBlock},
    basicstyle=\ttfamily\color{BlackText},
    keywords={
    },
    keywordstyle=\color{PurpleKeyword},
    ndkeywords={
        int, byte, return, new, class, public, private, static
    },
    ndkeywordstyle=\color{RedTypename},
    comment=[l][\color{cyan}\slshape]{//@},
    morecomment=[s][\color{cyan}\slshape]{/*}{*/},
    morecomment=[l][\color{GoldDocumentation}\slshape]{///},
    morecomment=[s][\color{GoldDocumentation}\slshape]{/*!}{*/},
    morecomment=[l][\color{GoldDocumentation}\slshape]{//!},
    morecomment=[s][\color{RedTypename}]{\#![}{]},
    morecomment=[s][\color{RedTypename}]{\#[}{]},
    morecomment=[l][\color{GrayComment}\slshape]{// a},
    stringstyle=\color{GreenString},
    string=[b]",
    escapeinside={(*}{*)}
}
\title{Predicate Subtypes in VerCors}
\author{Tycho Dubbeling \qquad
  Marieke Huisman \qquad \"Omer {\c S}akar
  \institute{University of Twente, Netherlands}
\email{t.b.dubbeling@student.utwente.nl \quad \{m.huisman,o.f.o.sakar\}@utwente.nl }
}

\def\titlerunning{Predicate Subtypes in VerCors}
\def\authorrunning{Dubbeling, Huisman \& {\c S}akar}
\begin{document}
\maketitle

\begin{abstract}

 Predicate subtypes provide an attractive mechanism to specify range
 constraints on variable declarations. This paper discusses how we add
 support for predicate subtypes to the VerCors program verifier. Our approach
automatically  generates  appropriate specifications from predicate
subtype declarations. It provides support to easily
combine multiple subtypes for a single variable declaration. Moreover, in order to use predicate
subtypes for overflow checking, a special strict mode is introduced,
where every subexpression also has to stay within the declared
subtype. A prototype implementation is integrated into the VerCors verifier.

  


 \end{abstract}

\section{Introduction}\label{sec:intro}

When reasoning about program behaviour, often constraints on program variables have to be specified, e.g., that the value of the variable is always  within a specific range of its specified type. One way of specifying such constraints is by using  predicate subtypes. A predicate subtype of a type \(X\) is defined as: all elements of \(X\) that satisfy a predicate \(P \colon X \rightarrow \mathsf{bool}\), i.e., formally \(\{x \colon X \mid P(x)\}\). In this paper, we discuss how we can support this way of specifying constraints on the values of a variable by adding support for predicate subtypes to the specification language of the VerCors~\cite{ArmborstBHHRST24} program verifier. VerCors is a deductive program verifier aimed in particular at the verification of concurrent programs. It uses a contract-based approach to specify the behaviour of a program. 

To support predicate subtypes, we allow user-defined predicates to be used as type modifers in our VerCors extension. An example of the usage of predicate subtypes is that a user may want to specify that the divisor of a division is not zero. The user can define a predicate \lstinline{NonZero} over integers where the integer does not equal zero, and then they can add this predicate as a type modifier to all the relevant variable declarations that are used as a divisor. Typical use cases for predicate subtypes are:
\begin{itemize}[noitemsep]
\item to automatically generate assertions in contexts where the same assertion needs to hold continuously throughout the program, and
\item to communicate information through the type system that may not be possible to encode in the type system of the programming language otherwise.
\end{itemize}
As type checking with predicate subtypes is undecidable, we need to do this outside the type checking system, by generating appropriate checks that ensure that the subtypes are respected. We do this by translating all predicate subtype declarations and usages into standard VerCors specifications, i.e., we generate extra pre- and postconditions and intermediate assertions. Note that the idea to use predicate subtypes is not new, but the advantage of our work is that generating appropriate specifications that capture that variables are within a specific range is fully automated, and compared to for example the support for subtypes in Dafny~\cite[\S 5.7]{DafnySub}, we have a bit more flexibility in what can be expressed (see Section~\ref{sec:related} for more details).

This paper discusses our implementation for predicate subtypes in VerCors. We discuss the general encoding principles, i.e., how subtype annotations are translated into first-order logic VerCors annotations, as well as how predicates can be combined into more complex subtypes.

One use case of predicate subtypes is to support reasoning about bounded integer domains (see e.g., \cite{prustiI32}). In VerCors, all numerical types, e.g., \lstinline{int} or \lstinline{byte}, are encoded as a mathematical integer type. However, this does not take into account that operations on them can overflow. As a result, we might miss possible program behaviours, as well as undefined behaviours or crashes. 
For example, given a C unsigned integer \lstinline{x} (i.e., an integer with a non-negative value) the program \lstinline{x = 0; x = x-1;} will overflow. This overflow is missed if unsigned integers are abstracted into mathematical integers. Integer overflows may also occur when typecasting to an integer type with stricter bounds. This paper shows how predicate subtypes also can be used to reason about such potential overflows. In particular, as overflows can also occur in intermediate expressions we introduce the notion of strict subtypes, which have to be respected by any subexpression, and which can therefore be used to check for the absence of overflows.


The remainder of this paper is organised as follows.
Section~\ref{sec:background} gives a brief introduction to VerCors. Section~\ref{sec:encoding} discusses how we encode predicate subtypes into VerCors, while Section~\ref{sec:strictness} shows how we can extend this to strict subtype checking for identifying overflows. Section~\ref{sec:implem} describes how we have implemented support for predicate subtypes in VerCors. Finally, Sections~\ref{sec:related} and \ref{sec:concl} discussed related work, conclusions and future work.

\section{Background}\label{sec:background}

VerCors\footnote{See \url{https://utwente.nl/vercors}} is a software verification tool that can reason about multiple (concurrent) programming languages, such as Java and C~\cite{ArmborstBHHRST24}. VerCors also supports its own verification language called Prototypal Verification Language (PVL). VerCors uses a contract-based approach to specify the expected behavior of a program, i.e., 
functions may be annotated with preconditions (\textbf{requires} clauses) and postconditions (\textbf{ensures} clauses). Specification properties are written in permission-based separation logic~\cite{AmighiHHH15}, an extension of Hoare logic~\cite{Hoare69} which extends first-order logic properties with operators from separation logic~\cite{Bornat05} and fractional permissions~\cite{Boyland03}. 

The architecture of VerCors is similar to a compiler's architecture. The input language is translated into an abstract syntax tree (AST). This AST gets translated through a series of passes into the format of the intermediate representation verifier Viper~\cite{EilersSSM25}, the backend of VerCors. For example, the Viper language does not support classes, thus, a pass within VerCors will translate Java-like classes into a representation with methods. The predicate subtypes in this work are supported by translating the predicate subtype specifications into (equivalent) specifications in VerCors' current specification language (see Section~\ref{sec:implem} for the implementation details).


\begin{figure}[t]
\begin{lstlisting}[language=JavaCors]
/*@	requires A != null;
	requires (\forall* int j; 0 <= j && j < A.length; Perm(A[j], write));
        ensures (\forall* int j; 0 <= j && j < A.length; Perm(A[j], write));
	ensures (\forall int j; 0 <= j && j < A.length; A[j] == 0); @*/
	void clearPar(int[] A) {
		par (int tid = 0 .. A.length)
		/*@	requires Perm(A[tid], write);
			ensures Perm(A[tid], write);
			ensures A[tid] == 0; @*/
		{ 
			A[tid] = 0;
		}
	} \end{lstlisting}
      \caption{Example VerCors PVL program and specification}\label{fig:vercors}
\end{figure}

Figure~\ref{fig:vercors} gives an example of a specification in VerCors. Method \lstinline{clearPar} sets all elements of an array \lstinline{A} to 0, and it does this in parallel (in a \texttt{par} block). The top level method specification indicates that the array is non-null, and that the thread that invokes the \lstinline{clearPar} method has the right to write to all elements of the array (written by special permission annotations \lstinline{Perm(A[j], write)}, see~\cite{AmighiHHH15} for more details. In the examples below, we leave out permission annotations, to focus on the subtype specifications). It also states that afterwards, all elements of the array will be put to 0. To verify the method, we specify for the parallel block the permissions that are given to each parallel thread, as well as the effect of each parallel thread.

\section{Predicate Subtypes}\label{sec:encoding}

This section describes how we encode predicate subtypes
into standard contracts. We first
describe the overall approach, and then illustrate this idea on several
examples. Finally, we show how multiple predicates can be combined
into a single predicate subtype.

\subsection{Encoding}

The goal of our encoding is to make sure that
a subtyped variable or expression always fulfills the specified
predicate subtype, at any point that it may be accessed. We do this by
creating appropriate pre- and postconditions for parameters and method
results, and by generating assertions whenever subtyped variables
are updated. More precisely:
\begin{itemize}[noitemsep]
    \item Subtypes on method input parameters are encoded by 
      preconditions for the method that assert that the subtyped parameters fulfill the relevant predicates.
    \item Subtypes on the return type of a method are encoded by
      postconditions for the method that assert that the return
      value of the function satisfies the subtype predicates.
    \item Expressions or statements that assign a value to a subtyped
      variable give rise to assertions that are added after the
      assignment, which assert that the new value of the variable
      satisfies the subtype predicates. 
    \end{itemize}

    We have chosen to generate assertions after each variable assignment
    because of ease of implementation in VerCors. An alternative
    approach would have been to add an assertion before the assignment that
    checks the result of the assignment expression. However, this approach 
    would entail that the right-hand side of the assignment needs to be 
    side-effect free, else the side-effect would occur twice, once in the 
    assertion and once in the assignment. The current choice made the 
    implementation easier, since the check is applied on the result of 
    the operation, after the possible side effects.
    
    Note that this list is not necessarily exhaustive: if a
    programming language has a different way of mutating values, this should 
    also be taken into consideration and appropriate assertions
    should be generated. Also type casts can be handled in this way,
    for any casts from \(A\) to \(B\), where \(B\) is a subtype, we
    can generate an assertion that the casted value should satisfy the
    predicate for \(B\).
    
    With this encoding, it is possible to specify that a
    variable should satisfy multiple predicates: this simple
    gives rise to multiple pre- or post-conditions, or
    assertions. More advanced approaches for combining predicate
    subtypes are discussed in Section~\ref{sec:multiple}.


\begin{figure}[t]
\begin{lstlisting}[language=JavaCors]
public class SubtypingExample {
    /*@ subtype NonZero(int x)() = x != 0;
        subtype len2(int[] xs)() = xs.length==2;
        subtype len3(int[] xs)() = xs.length==3;
        subtype Byte(int x)() = x >= -128 && x <= 127; 
        subtype NonNull(int[] x)() = x != null;
        subtype Index(int x)(int length) = x < length; @*/

    //@ ensures \result == x/y;
1   public static int division(int x, /*@ NonZero @*/ int y) {
        return x/y; }

2  private static void swap(/*@ nonNull len2 @*/ int[] xs) {
        int left = xs[0]; int right = xs[1];
        xs[0] = right; xs[1] = left; }

3   private static /*@ nonNull len3 @*/ int[] cross(
4        /*@ nonNull len3 @*/ int[] xs, /*@ nonNull len3 @*/ int[] ys) {
         int left = xs[1]*ys[2] - xs[2]*ys[1];
         int middle = xs[2]*ys[0] - xs[0]*ys[2];
         int right = xs[0]*ys[1] - xs[1]*ys[0];
         return new int[]{left,middle,right}; }

    public static void main(String[] args) {
5       /*@ Byte @*/ int result = division(3,5);
6       /*@ NonNull @*/ int[] array = new int[4];
7       /*@ Index(array.length) @*/ int index = 3;
        array[index] = result; }       
}

  
\end{lstlisting}
\caption{Example with predicate subtypes}\label{ex1}
\end{figure}

\begin{figure}[t]
\begin{lstlisting}[language=JavaCors]
1  /*@ requires y != 0;
       ensures \result == x/y; @*/
    public static int division(int x, int y) {
        return x/y; }

2   //@ requires xs!=null && xs.length==2;
    private static void swap(int[] xs) {
        int left = xs[0]; int right = xs[1];
        xs[0] = right; xs[1] = left; }

4   /*@ requires xs!=null && xs.length==3;
(*4*)       requires ys!=null && ys.length==3;
(*3*)       ensures \result!=null && \result.length==3; @*/
    private static int[] cross(int[] xs, int[] ys) {
        ..
    }     
      
    public static void main(String[] args) {
        int result = division(3,5);
5       //@ assert result >= -128 && result <= 127;
        int[] array = new int[4];
6       //@ assert array != null; 
        int index = 3;
7       //@ assert index < array.length.
        array[index] = result; }
}
\end{lstlisting}
\caption{Encoding of predicate subtypes in standard
  specifications}\label{ex1:encoded}
\end{figure}

\subsection{Example}
To illustrate our approach, consider the example in
Figure~\ref{ex1}\footnote{Where permission specifications are left out.}. 
It defines a class \lstinline{SubTypingExample} and four methods, namely 
\lstinline{division} to divide two integers, \lstinline{swap} to swap 
elements of an array with size 2, \lstinline{cross} to calculate the cross 
product of two arrays with size 3 and the \lstinline{main} method to call 
the methods above.

At the start of the class, it defines predicate subtypes
\lstinline{NonZero}, \lstinline{len2}, denoting that an array has
length 2, \lstinline{len3}, \lstinline{Byte}, \lstinline{NonNull}, and 
\lstinline{Index}, denoting that a value should be a valid
index in an array. 

The syntax to declare a predicate subtype is as follows:
\begin{lstlisting}[language=JavaCors]
//@ subtype <subtypeName>(<arg>)(<argList>) = <body>;
\end{lstlisting}
where the body of a subtype definition must be a boolean expression.

Notice that the predicate \lstinline{Index} has a
second parameter, thus this predicate provides an abstraction over a
range of possible subtypes. 
Notice that the
predicates \lstinline{len2} and \lstinline{len3} could have been written with such a second
parameter as well.

At any point that a type annotation occurs, a subtype may be specified\footnote{The 
\textbf{void} type in Java cannot be subtyped, since it does not behave like a proper type.}.
In the method \lstinline{division}, we specify that its
second parameter should be non-zero, while in the method \lstinline{swap},
we specify that its parameter should be a non-null array of
length 2. In the method \lstinline{main},
we declare several variables with a predicate
subtype. Figure~\ref{ex1:encoded} shows the same methods, where the predicate 
subtypes have been rewritten (automatically) to use only standard contract specifications 
using \lstinline{requires}, \lstinline{ensures} and 
\lstinline{assert} clauses. 
We numbered the lines with subtype usage in Figure~\ref{ex1}, and 
their corresponding specifications in Figure~\ref{ex1:encoded}.

   \subsection{Combining Multiple Subtypes}\label{sec:multiple}
    As illustrated in Figure~\ref{ex1}, a type declaration can be
    annotated with     multiple subtypes, e.g., the parameter of the \lstinline{swap} method. 
    In this case, it means that the type has to satisfy both predicates, i.e., a conjunction.
    However, sometimes it can also be useful to combine subtypes in other ways, so subtype
    definitions can be reused, \emph{e.g.}, via disjunction, implication or negation. We also
    provide syntax for this (in order of how strongly these operators bind):
    \begin{itemize}[noitemsep]
      \item  Disjunction: $\langle subtype \rangle | \langle subtype
        \rangle$.
        \item Implication: $\langle subtype \rangle ==> \langle
          subtype \rangle$.
          \item Conjunction: $\langle subtype \rangle \; \langle
            subtype \rangle$.
          \item Negation:  $! \langle subtype \rangle$.
          \end{itemize}
A toy example involving all operators is given in
Figure~\ref{fig:boolean}. The subtypes here express that the value of the \lstinline{arr} variable may only be either null, an array of length 10 or an array of length 3 (provided the array is not null). The value of \lstinline{x} must be both greater or equal to 0 and not smaller or equal than 0. 

\begin{figure}[t]
\begin{lstlisting}[language=JavaCors]
/*@ isNull | length(10) | nonNull ==> length(3) @*/ int[] arr;
arr = new int[10];
arr = new int[3];
/*@ nat !(negOrZero) @*/ int x = 1;
x = 0; // fails
\end{lstlisting}
\caption{Toy example illustrating the use of multiple ways to combine
  subtypes}\label{fig:boolean}
\end{figure}

\section{Subtypes and Overflow Checking}\label{sec:strictness}

A common use case for predicate subtypes (and an initial
motivation for our work) is to check for the absence of
overflows. When checking for overflows, we need to ensure that also
intermediate expressions stay within the bounds of the
subtype. Therefore, subtypes may be declared with the keyword
\lstinline{strict} in front of it.  For example, the following would
fail to verify:
\begin{lstlisting}[language=JavaCors]
/*@ strict range(0,128) @*/ int x = 0;
int y = (x-2)+2; // fails
\end{lstlisting}
because the intermediate expression x - 2 does not stay within the
subtype \lstinline{range(0, 128)}.
This is despite the final return value of the entire expression satisfying the subtype predicate. 

To facilitate the checking of overflows, our implementation in VerCors
provides a  "--strict-arithmetic" option flag, which  will convert any
\lstinline{int} type into a strictly subtyped \lstinline{int}, using the following subtype definition: 
\begin{lstlisting}[language=JavaCors]
/*@ subtype strictIntGenerated(int X)() = 
            -2147483648 <= X && X <= 2147483647; @*/
\end{lstlisting}
This subtype precisely checks whether a value is within the bounds of a Java \lstinline{int} type. 
As a result, the strict-arithmetic option flag adds overflow checks for all expressions involving Java's \lstinline{int}.

\section{Implementation}\label{sec:implem}

We implemented prototype support for predicate subtypes in VerCors.
The predicate subtyping is supported for all types, including user-defined types.
The process VerCors goes through during verification can be described
in 5 major steps. (1)~VerCors parses the program. The grammar
VerCors parses is built out of a specification grammar and a grammar
for the language it needs to parse. (2)~We
build a syntax tree out of the parsed nodes. During
this step, VerCors may already produce an error if it detects
an unsupported feature in the parsed nodes. (3)~VerCors
tries to resolve references and types within the parsed
program. (4)~During the transformation step, VerCors will
apply a list of rewriters to reduce the syntax tree into a form that
is accepted by the back-end. (5)~The
rewritten syntax tree is given to the Viper~\cite{EilersSSM25} back-end for
verification. 

We have added subtyping annotations to the Java parser for VerCors and
subtype declarations to the specification parser for
VerCors. During the parsing stage, subtype annotations get interpreted as
expressions. The boolean operators between subtypes are directly
interpreted as the \lstinline{Expr} nodes for the corresponding
operator. This expression gets stored within a Type node
called \lstinline{TSubtype}, which also stores whether the annotation was
declared strict as a separate parameter. 

Furthermore, we have added several rewrite steps to the
transformation phase of VerCors. Currently, the prototype
implementation supports the  rewrites for parameters and method
results and generates an assert at variable assignments. It also
supports the use of predicates with multiple parameters, such as the
\lstinline{Index} predicate in Figure~\ref{ex1}. 

The rewriting is done in three steps. First, there is the transformation step that turns
\lstinline{int}'s into subtyped integers whenever the strict-arithmetic
flag has been set. This transformation step always creates a new subtype declaration
and adds it to the global scope of the provided program. The subtype
declaration uses hard coded integer values to set the integer bounds
for the subtype.
Then, there is the transformation step that
rewrites subtypes of subtypes to handle the nesting. When a subtype
declaration uses another subtype, the predicate corresponding to the annotation subtype gets conjugated to the subtype's body. For example: 
\begin{lstlisting}[language=JavaCors]
//@ subtype pos(subtype<int,nat | range(0,128)> x)() = x >= 1;
\end{lstlisting}
gets rewritten to:
\begin{lstlisting}[language=JavaCors]
/*@ subtype pos(subtype<int,nat | range(0,128)> x)() = x >= 1
                && (x >= 0 | (x >= 0 && x <= 128));
@*/
\end{lstlisting}
Finally, we converts all subtype annotations into the
appropriate verification checks.

As explained before, the predicate subtypes are translated into
equivalent specifications that are part of VerCors' current
specification language (i.e., functions, pre-/postconditions). The
translation of predicate types is in principal safe, since the
predicate subtypes are translated into functions, which are well
supported in VerCors. Any additional transformations required, e.g.,
on the generated functions, are already part of the series of passes
present in VerCors, such as type checking and the translation into the Viper format. 

In our encoding, we decided  to inline the subtype
predicates. If we had kept the predicate names in the specifications,
they would have to be explicitly unfolded\footnote{I.e., replacing the
  predicate name by its body}, similar to how other predicates are
handled in
VerCors. We made this
    choice because the subtyping system may generate a lot of proof
    obligations for subtyped variables, and we believe inlining them keeps the extra
    proof effort reasonable.
    
As mentioned above, our implementation provides an option flag that allows for checking of
overflows. This makes it possible to first verify a program assuming
mathematical numbers, and then verify that it does not have any overflows.

\section{Related Work}\label{sec:related}

Our idea are quite close in spirit to the \lstinline{subsets} and \lstinline{newtypes} mechanisms in Dafny~\cite[\S 5.6.3 and \S 5.7]{DafnySub} (but developed independently). 
Newtypes are always strict, i.e., also all subexpressions have to satisfy the predicate. 
Meanwhile, subset types behave like non-strict subtypes.
Dafny also generates a condition to prove that a subtype is non-empty. Dafny does not support parametrised subtypes, that are instantiated later. For example, the following predicate subtype cannot be defined in Dafny. 
\begin{lstlisting}
//@ subtype range(int x)(int lower, int upper) = lower >= x && x >= upper;
/*@ range(0,128) @*/ int variableName = 0;
\end{lstlisting}

There also has been previous work on encoding bounded integers in program verifiers. In particular,  Prusti \cite{prusti}, Rust front-end   built on Viper, encodes 32-bits signed integers in Viper \cite{prustiI32} by introducing a special \lstinline{i32} predicate in the encoding, which is used when overflow checks are turned on. 
\begin{lstlisting}[language=Viper]
predicate i32(self: Ref) { 
  acc(self.val_int, write) && 
    -2147483648 <= self.val_int && 
    self.val_int <= 2147483647 
} 
\end{lstlisting}
This partially inspired and confirmed our ideas, but we further generalised it to allow arbitrary predicate subtypes to be specified. 

Also Frama-C~\cite{KosmatovPS24} provides support to provide the absence of integer overflows. This is done by the RTE plugin of Frama-C, which can
generate assertions checking for overflows at every downcast \cite{RTE}. Furthermore, it also can generate annotations for all arithmetic calculations that were performed. The user can choose themselves whether these annotations should be generated (and verified). Furthermore, the WP plugin~\cite{WP} for Frama-C supports two models for integer arithmetic: the machine integer model and the natural model. 
The machine integer model assumes that signed integers are not allowed to overflow, while unsigned integer arithmetic is allowed to overflow, and this will be verified.
The natural model uses mathematical operations on all integer types, and no bounds will be verified. We use a similar approach, as this allows us to control how many extra proof obligations need to be handled.

There also has been further previous work on implementing predicate subtyping. For example, the PVS verification system allows the user to specify predicate subtypes, and generates appropriate verification conditions (called type checking conditions (TCC's)) for them~\cite{subIEEE}.

\section{Conclusions and Future Work}\label{sec:concl}

Currently, there is support for predicate subtyping in the grammars used by VerCors for parsing Java and PVL. Furthermore, VerCors currently supports the disjunction, implication, conjunction and negation operators for combining subtypes as described in Section~\ref{sec:multiple}. There is also support for subtypes of subtypes. Strict subtype checking of expressions has been implemented for expressions built from Numeric Binary operators. Finally, an option flag for strict overflow checking for 32-bit integers has been added to VerCors.

There are further improvements that can be made to the current predicate subtyping implementation.
First, the strict arithmetic flag only supports overflow checking of signed 32-bit integers (Java's \lstinline{int}). 
The reason for this is that all Java integer types get converted to \lstinline{TInt} in the version of VerCors the extension was developed with. 
For the strict arithmetic flag to handle different sized integer types, the transformation steps for strict arithmetic need to distinguish the different Node Types. 
This requires distinction between mathematical integers and bounded integers in VerCors' AST and transformations.
Currently, VerCors has a bounded integer type defined, it is currently not used to distinguish between Java's different integer types.

The predicate subtyping implementation only supports Java and PVL. 
As such, adding predicate subtypes to the C front-end of VerCors is a natural follow-up future work. 
Eventually, it would be nice to have a standard library of predefined subtypes, with generic subtypes that could be used, as well as Java- or C-specific types, which would only be usable in their respective languages.

Another future work could be adding a \lstinline{NonNull} subtype whenever a \lstinline{@NonNull} Java annotation is present. This would avoid repeating information at the function interface. We could also add support for subtypes of variables declared in forall or exists predicates.

Finally, there is still room for improvements regarding the error messages. Currently, the predicate is transformed away, meaning that when an error occurs, the error message does not point to the predicate subtype. This can make it harder to understand the issue.

 \bibliographystyle{eptcs}
 \bibliography{generic}
\end{document}